\documentstyle[12pt,epsf,epsfig]{article}

\def\beq     {\begin{equation}} 
\def\eeq     {\end{equation}}  
\def\beqa    {\begin{eqnarray}}    
\def\eeqa    {\end{eqnarray}}   
\def\pom     {\mbox{{\scriptsize P}\hspace{-0.16cm}{\scriptsize I}}} 
\def\bigpom  {\mbox{ P\hspace{-0.22cm}I}} 
\def\shift   {\rule[-3mm]{0mm}{8mm}}

\begin{document}

\begin{titlepage}

\vspace*{0.1cm}   
\begin{flushright}  
DTP/96/10      \\
February  1996  \\   
\end{flushright}   
\vskip   1.2cm   

\begin{center}
{\Large\bf   
Diffractive  Heavy  Flavour  Production  \\[2mm]  
at the Tevatron  and the LHC } 
\vskip 1.cm 
{\large  M.~Heyssler}  
\vskip .3cm
{\it  Department  of Physics,  University of Durham \\ 
Durham DH1 3LE, England }\\

\vskip 1cm

\end{center}

\begin{abstract}

We give  predictions for diffractive  heavy flavour  production at the
Tevatron  and  the  LHC  in   leading--order   approximation.  In  the
framework  of these  studies  we use three  different  models  for the
partonic  structure of the Pomeron  recently  proposed by Stirling and
Kunszt.  These  Pomeron  models are, despite  being fitted to the same
diffractive  deep inelastic  HERA data, very different in their parton
content  and  taken  together  provide  a  powerful  tool to probe the
structure of the Pomeron.  All models satisfy GLAP  evolution and show
a significant  $Q^2$  dependence.  We give numerical  predictions  for
single  as  well as  double  diffractive  cross  sections  assuming  a
Donnachie--Landshoff--type Pomeron flux factor.

\end{abstract}

\vfill 

\end{titlepage}

\newpage


\section{Introduction}

Recent observations at HERA \cite{H194,ZEUS94} have provoked a renewed
interest  in an already  long  standing  concept,  the  concept of the
Pomeron in high energy  scattering.  This idea was first introduced by
Pomeranchuk in 1958  \cite{pomeranchuk58}  to explain the fact that in
the high energy limit the  measured  hadronic  cross  sections  remain
(approximately)  constant.  This asymptotic behaviour was explained in
terms of a {\em  Regge  trajectory}  $\alpha(|t|)$  with  the  Pomeron
trajectory having the intercept $\alpha(0)\approx 1$.

Present and future  generations of hadron--hadron  and  lepton--hadron
colliders may be able to actually  {\em  observe} the  Po\-mer\-on  in
diffractive  events.  We shall  characterise  an event as {\em  single
diffractive}  if one of the  colliding  hadrons  emits a Pomeron ${\rm
\bigpom}$  that  scatters off the other hadron.  A single  diffractive
reaction would therefore correspond to
\beq  \label{reaction}  
\mbox{\rm hadron~$i$} + \mbox{\rm  hadron~$j$}
\longrightarrow      
\mbox{\rm hadron~$i$} + X,     
\eeq
in  which  the   intact   hadron~$i$   emits  a  Pomeron   (hadron~$i$
$\rightarrow$  hadron~$i$ + $\bigpom$) which interacts with hadron~$j$
(hadron~$j$ + ${\rm\bigpom} \rightarrow X$) leading to the reaction of
Eq.~({\ref{reaction}).  Hadron~$i$ is detected in the final state with
a large  longitudinal  momentum  fraction.  Hard  diffractive  events,
events with a large  momentum  transfer $Q$ ($1/Q\ll  1$~fm), are also
characterised  by a second feature:  the absence of hadronic energy in
certain  angular  regions of the final state phase  space.  These {\em
rapidity  gaps}, reported by the ZEUS and H1  collaborations  at HERA,
are suggestive of a colour neutral  object, the Pomeron,  emitted from
the  incoming  proton.  Events  fulfilling  the  conditions  of  large
rapidity gaps and a (highly excited) hadron remnant in the final state
are called {\em single  diffractive} to distinguish them from those in
which  both  colliding  hadrons  remain  intact  as they  each  emit a
Pomeron,  the  so--called  {\em  double  diffractive}  events.  The H1
collaboration  reported  a  fraction  of about  6\% of deep  inelastic
scattering events being single diffractive \cite{H192}.  The HERA data
suggest that  ${\rm\bigpom}$  has a hard structure.  This  observation
has also been given support in an earlier UA8  experiment.  A partonic
structure of the Pomeron was first proposed by Ingelman and Schlein in
1985  \cite{ingelman85}  to calculate  high  $p_T$--jet  production in
single diffractive  dissociation.  The discovery of diffractive dijets
(high  $p_T$--jets) with $p_T>8$~GeV at the CERN  $p\bar{p}$--collider
in 1988 by the UA8 collaboration  \cite{UA8} fitted exactly with their
prediction.

In this paper we give  predictions  for single and double  diffractive
heavy   flavour   production  at  the  Tevatron   $p\bar{p}$--collider
($\sqrt{s}=   1.8$~TeV)   and   the   LHC   $pp$--collider,   with   a
centre--of--mass  energy  $\sqrt{s}=10-14~$~TeV.  By heavy  flavour we
mean  $Q\bar{Q}$  production  with $Q=c$  (charm), $b$ (bottom) or $t$
(top).  Even though  several  tests of  diffractive  $W$, $Z$ or Higgs
production   \cite{diffall,james96,bruni93,nachtmann90}  have  already
been proposed, we shall focus especially on $Q\bar{Q}$ states as their
cross  sections  turn out to be typically  larger and this makes their
observation,   especially  in  the  diffractive   context,  easier  in
principle.  We  shall  employ  different   aspects  all  deduced  from
collider physics phenomenology to create a solid model for diffractive
scattering via Pomeron exchange.

The emission of a Pomeron from a  fast--moving  hadron will be treated
in the framework of the  Donnachie--Landshoff  model  \cite{dola} of a
reggeised,  non--perturbative Pomeron.  In the literature this type of
Pomeron is called {\em soft} to discern it from the perturbative ({\em
hard})  Pomeron.  The  number  of  Pomerons  emitted  by a  hadron  is
described by the flux factor  $f_{{\rm\pom}/i}(x^{\rm\pom},|t|)$ where
$i$  stands  for  any  hadron.  The  longitudinal   momentum  fraction
$x^{\rm\pom}$ of the Pomeron and the $t$--channel  invariant  momentum
transfer  to  the  Pomeron  fully   determine   the  emission   factor
$f_{{\rm\pom}/i}(x^{\rm\pom},|t|)$.   As   will   be   discussed    in
Section~2, a Regge factor is imbedded in the flux factor  instead of a
Pomeron propagator.

The success of this  prediction has been exploited to expand the ideas
of the  parton  model to the  Pomeron.  The  Pomeron,  being a  colour
neutral object  carrying the quantum  numbers of the vacuum, should be
made    of    $q\bar{q}$    pairs    and/or    gluons.   Factorisation
\cite{ingelman93}   defines   the   diffractive   structure   function
$F_2^D(x,Q^2,x^{\rm\pom},|t|)$  as a product of the structure function
of the  Pomeron  $F_2^{\rm\pom}(x,Q^2)$  and  the  flux  factor,  i.e.
\beq       \label{factor1}        
F_2^D(x,Q^2,x^{\rm\pom},|t|) = F_2^{\rm\pom}(x,Q^2)  
\otimes   f_{{\rm\pom}/i}(x^{\rm\pom},|t|).  
\eeq
Measuring the diffractive  structure  function  $F_2^D$ and assuming a
particular flux (emission) factor\footnote{The flux factor can also be
extracted from experiment.}  $f_{{\rm\pom}/i}(x^{\rm\pom},|t|)$  gives
information on the Pomeron structure  function  $F_2^{\rm\pom}(x,Q^2)$
and hence on the parton content of the Pomeron.  Because factorisation
is assumed to give a universal  picture of the Pomeron  structure, the
HERA data on  $F_2^{\rm\pom}(x,Q^2)$  can be used to give  predictions
for  hard   diffractive   scattering  at   hadron--hadron   colliders.
Comparing these predictions with measurements allows the factorisation
hypothesis to be tested.

This leads back to the fact that $F_2^D$ has recently been measured at
HERA  \cite{HERA}  and thus provides new  information  on the partonic
structure of the Pomeron.  The data still suffer from large errors but
nevertheless allow a fit of possible parton  distributions  inside the
Pomeron.     This     has     been     done     for     example     in
Refs.~\cite{james96,models,tom95}.       In       particular,       in
Ref.~\cite{james96}  three Pomeron  models were proposed to fit recent
H1 data \cite{HERA}.  At this stage it has not been possible to decide
experimentally  whether the Pomeron is more gluonic or  quark--like --
deep  inelastic  scattering  (DIS) only gives  information on the {\em
quark} content.  Taking this deficiency into account, the three models
range from pure glue up to pure quark  content at the  starting  scale
$Q_0=2$~GeV.  While  the quark  content  in each of the  three  models
turns out to be rather similar, the gluon contents differ from soft to
hard.  No momentum  sum rule for the  distributions  is imposed and an
overall  normalisation factor is incorporated into the flux factor, as
described in  Ref.~\cite{tom95}.  A more  detailed  discussion  can be
found in Section~3.

The  aims of this  paper  are as  follows.  We want to  show  how  the
combination  of Regge  theory  and  factorisation  can be  applied  to
calculate  diffractive  cross  sections,  especially  in the  light of
recent  experimental  $F_2^D$ data.  The theory of this is reviewed in
Section~2.  So far the  theory  has  been  mostly  applied  to  single
diffractive  events.  The  reason  is that  double  diffractive  cross
sections at present and future  colliders  are generally  too small to
play  a  pivotal  or  even  measurable  role.  But  especially   after
experiences with top quark production and recent measurements of charm
and bottom quark total cross  sections at the Tevatron  \cite{CDF} and
elsewhere,  it is worth looking for heavy flavour  production  in both
single  as  well  as  double   diffractive   cross   sections.  Future
measurements at high energy hadron--hadron colliders together with our
predictions could be a powerful step towards a better understanding of
the Pomeron structure, in particular to reveal a gluonic,  quark--like
or  mixed  Pomeron.  We  shall  also  consider  the   next--generation
collider,  the LHC, for this  purpose.  With a centre--of--mass energy
$\sqrt{s}=10$~TeV,  we shall show that double  diffraction  will be no
longer  an  unmeasurable   effect  but  a  paradigm  for   visualising
Pomeron--Pomeron scattering.

We note that the  concept  of a  universal  factorisation  of  Pomeron
structure, in particular for  hadron--hadron  collisions, has recently
been questioned, see for example Ref.~\cite{soper95}.  Our predictions
and  further  experiments  at the  Tevatron  and the LHC might help to
clarify this question.

All numerical  predictions  for the Tevatron and the LHC are presented
in Section~4.  The results are critically  discussed and the different
models  are  carefully  discerned.  Both  absolute  values  as well as
single and double diffractive ratios are shown.  Section~5  summarises
the results, underlines the most important features and discusses open
problems.


\section{Kinematics and Cross Sections}

In this paper we  concentrate on the single  $\sigma^{SD}$  and double
diffractive  $\sigma^{DD}$  cross sections for the production of heavy
flavour  $Q\bar{Q}$  states.  What do we  mean  by {\em  diffractive}?
First let us consider the case of single  diffraction, where a Pomeron
with  momentum  $k_i^{\rm\pom}$  is  emitted  by one of the  colliding
hadrons  as  shown  in   Fig.~1a.  The   definition   of  {\em  single
diffractive}  means:  one  hadron,  emitting  the  Pomeron,  is  being
detected (at least in principle) in the final state.  The other hadron
scatters  off  the  emitted  Pomeron.  A  typical  single  diffractive
reaction for our purpose at the LHC would read:  $p+p  \longrightarrow
p + Q\bar{Q} + X$.  The Pomeron kinematical  variable  $x_i^{\rm\pom}$
is    defined    as:    $x_i^{\rm\pom}=s_j^{\rm\pom}/s_{ij}$,    where
$\sqrt{s_j^{\rm\pom}}$   is  the   centre--of--mass   energy   in  the
Pomeron--hadron~$j$    system   and    $\sqrt{s_{ij}}=\sqrt{s}$    the
centre--of--mass energy in the hadron~$i$--hadron~$j$ system.  Instead
of   the   gluon--   or   (anti)quark--distributions   of   hadron~$i$
($f_{g,q/i}(x_i,Q^2)$),   the  parton  distributions  of  the  Pomeron
($f_{g,q/{\rm\pom}}(x_i/x_i^{\rm\pom},Q^2)$     with     $x_i     \leq
x_i^{\rm\pom}$)  emitted by hadron~$i$  define the single  diffractive
heavy quark cross sections assuming a partonic picture of the Pomeron.
We have therefore to substitute
\beq     \label{substitution}      
f_{g,q/i}(x_i,Q^2) \Rightarrow
\int\limits_{|t_i|} d|t_i| \int\limits_{x_i^{\rm\pom}}
\frac{dx_i^{\rm\pom}}{x_i^{\rm\pom}}\,
f_{g,q/{\rm\pom}}(x_i/x_i^{\rm\pom},Q^2)\otimes
f_{{\rm\pom}/i}(x_i^{\rm\pom},|t_i|).                             
\eeq
The Pomeron parton distribution $f_{g,q/{\rm\pom}}(x_i^{\rm\pom},Q^2)$
is weighted  by a flux  factor  $f_{{\rm\pom}/i}(x_i^{\rm\pom},|t_i|)$
which gives a measure of the number of Pomerons emitted by hadron~$i$.
It is  therefore a function  of the energy  loss of the proton  $t_i =
(P_i-P_i')^2$  due to  Pomeron  emission  as  well as of the  momentum
fraction  $x_i^{\rm\pom}$  carried  by the  Pomeron.  Its  form can be
modelled by~\cite{dola}
\beq     \label{flux}      
f_{{\rm\pom}/i}(x_i^{\rm\pom},|t_i|)  =
\frac{9\beta^2}{4\pi^2}    
\left[ F_1(|t_i|) \right]^2    
\left( x_i^{\rm\pom}\right)^{1-2\alpha(|t_i|)}.                          
\eeq
The  coupling  of the  Pomeron to the proton is  assumed to be $3\beta
F_1(|t_i|)$ with $\beta=1.8~{\rm GeV}^{-2}$.  The factor 3 corresponds
to the number of valence  quarks inside the hadron.  This is discussed
and supported in \cite{dola}.  In fact it should couple in a pointlike
form to single quarks.  From this point of view the effective  Pomeron
structure function should contain quark--antiquark  contributions.  An
analogy with the photon structure  function emerges from this picture.
The elastic form factor of the proton  $F_1(|t_i|)$ is  experimentally
determined and well parametrised by \cite{dola}
\beq  \label{form}  
F_1(|t_i|) = \frac{  4M_p^2 + 2.8|t_i| }{ 4M_p^2 + |t_i| } 
\left( 1 + \frac{  |t_i| }{  0.7~{\rm  GeV}^2 }  \right)^{-2},
\eeq
where  $M_p$  is the  proton  mass.  Instead  of  propagators  for the
Pomeron,     one     inserts     the     Regge     factor      $\left(
x_i^{\rm\pom}\right)^{1-2\alpha(|t_i|)}$  with  the  Regge  trajectory
$\alpha(|t_i|)  = 1 + \epsilon -  \alpha'|t_i|$  with  $\epsilon$  and
$\alpha'$  both  deduced  experimentally.  Latest  experiments  at the
Tevatron   \cite{CDF94}  showed  a  slight  $\sqrt{s}$  dependence  of
$\epsilon$  as  long as  $\alpha'$  is  fixed.  The CDF  collaboration
studied single diffractive events $p + \bar{p}  \rightarrow  \bar{p} +
X$ and obtained from a data fit:
\begin{eqnarray*}  
\epsilon = 0.121\pm  0.011  &{\rm at}&  \sqrt{s} = 546~{\rm  GeV}, \\  
\epsilon = 0.103\pm  0.017  &{\rm at}&  \sqrt{s} =1800~{\rm  GeV}.
\end{eqnarray*}
They  conclude that large  screening  effects have to be introduced to
save the  traditional  Pomeron  model.  An analysis at $\sqrt{s} = 20,
546$ and $1800$~GeV showed a significant  $\sigma^{SD}\sim  s^{0.030}$
dependence for the single diffractive cross section.

In  practice,  $\epsilon$,  introduced  as an {\em  effective  power},
measures   the   deviation   of   the   differential   cross   section
$d\sigma^{SD}/dM^2$  from a pure  $1/M^2$  dependence.  On  the  other
hand,  taking  only the  triple--Pomeron  diagram as  contribution  to
diffractive   dissociation  into  account  and  neglecting   screening
effects,  $1+\epsilon$  can  be  interpreted  as  the  intercept  of a
supercritical Pomeron \cite{white95}.

Also deduced  experimentally  by CDF was the sensitivity to $\alpha'$.
A change in  $\alpha'$ by  $\delta\alpha'  = \pm  0.1~{\rm  GeV}^{-2}$
results  in  a  change  in  $\sigma^{SD}$  of  only  $\pm  0.1$\%  and
$\epsilon$  changes at the same time by  $\delta\epsilon = \pm 0.011$.
Neglecting any $s$  dependence, we can fix $\epsilon$ and $\alpha'$ to
the    standard    values     \cite{dola}     $\epsilon=0.085$     and
$\alpha'=0.25$~GeV$^{-2}$.

According to \cite{dola} all these  relations  should only be valid if
we limit the Pomeron  momentum  fraction to  $x_i^{\rm\pom}\leq  0.1$.
Thus only a maximum  energy loss of 10\% for each  hadron is  allowed.
But assuming the final state proton is not detected,  there will be no
energy cut for $t$, although the flux factor  decreases very fast with
$|t|$.

The flux factor  plays a pivotal  role because it directly  yields the
energy  dependent  emission  rate  of the  Pomerons  and is  therefore
extensively   discussed   in  the   literature.  Bruni  and   Ingelman
calculated the diffractive $W$ production  rates at the Tevatron using
a constant  Pomeron--proton  cross section of $2.3$~mb,  obtained from
Regge  analysis  but with a  non--Regge  parametrisation  for the flux
factor \cite{bruni93}
\beq     \label{bruni}      
f_{{\rm\pom}/i}(x_i^{\rm\pom},|t_i|) = \frac{1}{x_i^{\rm\pom}}  
\big( 6.38  e^{-8|t_i|}  + 0.424  e^{-3|t_i|} \big) \frac{1}{2.3},
\eeq
fitted from single  diffractive  data.  Throughout this work, however,
we shall use the  Donnachie--Landshoff  Pomeron flux  factor\footnote{
Goulianos  has  recently   argued   \cite{goulianos95}   that  only  a
normalisation  of the flux  factor to {\em one  Pomeron  per  nucleon}
might yield physical results.}  of Eq.~(\ref{flux}).

We shall also consider  double  diffractive  events.  The procedure is
analogous  to the  single  diffractive  scenario  and is  sketched  in
Fig.~1b.  Now  {\em  both}  hadrons   undergo  Pomeron   emission  and
therefore all hadronic parton distributions are replaced by the parton
distributions    of   the    Pomeron    via   the   rule    given   in
Eq.~(\ref{substitution}).   However,   the   basic    formulation   of
diffractive  scattering  was intended for  reactions  where one hadron
scatters  diffractively and one hadron is highly excited.  The crucial
point is that the single diffractive cross section of a hadron--hadron
collision is assumed to factorise into the total Pomeron--hadron cross
section and the  Pomeron  flux  factor  \cite{ingelman85}.  The single
diffractive event may then be written as
\beq  \label{factor}  
\frac{d\sigma^{SD}(p + p(\bar{p})  \rightarrow
p(\bar{p}) + Q\bar{Q} + X)}    
{dx_i^{\rm\pom}d|t_i|} = f_{{\rm\pom}/i}(x_i^{\rm\pom},|t_i|)
\otimes \sigma^T({\rm\bigpom}+p(\bar{p}) \rightarrow  Q\bar{Q}  +  X).  
\eeq
Eq.~(\ref{factor})   is  the  analogue  of   Eq.~(\ref{substitution}).
Factorisation has also been applied to double diffraction, for example
predictions  for  Higgs  \cite{nachtmann90}  or $W$  boson  production
\cite{james96,bruni93} at the Tevatron and/or the LHC.  Both colliding
hadrons  can in  principle  be detected in the final  state.  Again, a
typical reaction at the LHC would look like:  $p+p \longrightarrow p+p
+ Q\bar{Q} + X$.  Double  diffractive events thus are characterised by
two  quasi--elastic  protons with  rapidity  gaps between them and the
central heavy flavour products.

Using  the  substitution  given  in  Eq.~(\ref{substitution})  we  can
proceed in complete analogy to the single diffractive case and finally
write  down  the   expressions   for  the  total,  single  and  double
diffractive     cross    sections    for     $Q\bar{Q}$     production
\beqa        \sigma^T(s,m_Q)       &=&        \int\limits_{x_1=\tau}^1
\int\limits_{x_2=\frac{\tau}{x_1}}^1      \;     dx_1dx_2      \Big[\,
f_{q/1}(x_1,Q^2)                                f_{q/2}(x_2,Q^2)\times
\hat{\sigma}_{q\bar{q}}(\hat{s},m_Q,\alpha_S(Q^2))  \nonumber  \\  &+&
f_{g/1}(x_1,Q^2)                                f_{g/2}(x_2,Q^2)\times
\hat{\sigma}_{gg}(\hat{s},m_Q,\alpha_S(Q^2))\,                \Big],\\
\sigma^{SD}(s,m_Q)        &=&       \int       dx_1\int       dx_2\int
\frac{dx_1^{\rm\pom}}{x_1^{\rm\pom}}         \int        d|t_1|\otimes
f_{{\rm\pom}/1}(x_1^{\rm\pom},|t_1|)    \nonumber\\   &\times&   \Big[
f_{q/{\rm\pom}}(x_1/x_1^{\rm\pom},Q^2)                f_{q/2}(x_2,Q^2)
\times\hat{\sigma}_{q\bar{q}}(\hat{s},m_Q,\alpha_S(Q^2))   \nonumber\\
&+&      f_{g/{\rm\pom}}(x_1/x_1^{\rm\pom},Q^2)       f_{g/2}(x_2,Q^2)
\times\hat{\sigma}_{gg}(\hat{s},m_Q,\alpha_S(Q^2))\;\Big]  \nonumber\\
&+&                     (1\rightleftharpoons                     2),\\
\sigma^{DD}(s,m_Q)        &=&       \int       dx_1\int       dx_2\int
\frac{dx_1^{\rm\pom}}{x_1^{\rm\pom}}                              \int
\frac{dx_2^{\rm\pom}}{x_2^{\rm\pom}}\int  d|t_1| \int d|t_2| {\otimes}
f_{{\rm\pom}/1}(x_1^{\rm\pom},|t_1|)
f_{{\rm\pom}/2}(x_2^{\rm\pom},|t_2|)     \nonumber\\     &\times&\Big[
f_{q/{\rm\pom}}(x_1/x_1^{\rm\pom},Q^2)
f_{q/{\rm\pom}}(x_2/x_2^{\rm\pom},Q^2)
\times\hat{\sigma}_{q\bar{q}}(\hat{s},m_Q,\alpha_S(Q^2))   \nonumber\\
&+&                             f_{g/{\rm\pom}}(x_1/x_1^{\rm\pom},Q^2)
f_{g/{\rm\pom}}(x_2/x_2^{\rm\pom},Q^2)
\times\hat{\sigma}_{gg}(\hat{s},m_Q,\alpha_S(Q^2))\Big].         \eeqa
Of  course  the  threshold  condition  $\hat{s}=x_1x_2s\geq  \tau  s =
4m^2_Q$ for the production of a $Q\bar{Q}$ pair of mass $2m_Q$ must be
fulfilled   and  the  cut--off  in  the  Pomeron   spectrum   $x_i\leq
x_i^{\rm\pom}\leq  0.1$ taken into account.  With  $\hat{s}$ we define
the Mandelstam variable of the two contributing subprocesses $q\bar{q}
\rightarrow  Q\bar{Q}$  and $gg  \rightarrow  Q\bar{Q}$  and  $\tau  =
4m^2_Q/s$ regulates the threshold condition  automatically as function
of the integration  boundaries\footnote{For  the sake of simplicity we
did  not  explicitly   write  down  the  integration   boundaries  for
$\sigma^{SD}$ and  $\sigma^{DD}$.  They can easily be deduced from the
two    conditions:   (1)    $x_1x_2\geq\tau$    and    (2)    $x_i\leq
x_i^{\rm\pom}\leq  0.1$.}. We restrict ourselves to the leading--order
contributions.  The Feynman  diagrams of this Born  approximation  are
shown   in   Fig.~2.    A   discussion   of   next--to--leading--order
non--diffractive   heavy   flavour   production   can  be   found   in
\cite{nason92}.

We only cite the  expressions  for the  subprocess  cross  sections in
leading--order approximation. With  $\rho=4m^2_Q/{\hat{s}}$  they read
\beqa  \label{subcross}   \hat{\sigma}_{gg}(\hat{s},m_Q,\alpha_S(Q^2))
&=&         \frac{1}{48}\frac{\pi\alpha^2(Q^2)}{\hat{s}}         \Big(
-(112+124\rho)\sqrt{1-\rho}  \nonumber  \\ &+& (16 + 16\rho +  \rho^2)
\ln\left(  \frac{1+\sqrt{1-\rho}}{1-\sqrt{1-\rho}}  \right)  \Big), \\
\hat{\sigma}_{q\bar          q}(\hat{s},m_Q,\alpha_S(Q^2))         &=&
\frac{16}{27}\frac{\pi\alpha^2(Q^2)}{\hat{s}}  \left(  1+\rho  \right)
\sqrt{1-\rho},                                                   \eeqa
where  the  renormalisation  scale  is  taken  to  be  the  subprocess
collision energy, $Q^2 = \hat{s}=x_1x_2s$.  For an analytic derivation
of Eqs.~(11) and (12) using standard techniques we refer the reader to
\cite{jones78,gluck78,combridge79}.

One  might  however  argue  that  for  the  $gg\rightarrow   Q\bar{Q}$
subprocess the $\hat{t}$-- and $\hat{u}$--channel contributions should
dominate \cite{gluck78}, in which case a choice of $Q^2 = \frac{1}{2}$
$\left(  (m^2_Q-\hat{t})  +  (m^2_Q-\hat{u})\right)$   =  $\frac{1}{2}
\hat{s}$ might be more  reasonable.  In fact the parton  distributions
{\em are}  affected  by the choice of $Q^2$ in the  framework  of GLAP
evolution.  But as in  \cite{combridge79}  we could  {\em  not} find a
significant  sensitivity  of our results to this choice  compared to a
general $Q^2=\hat{s}$ for both subprocesses.

A still unsolved  problem is the structure of the Pomeron.  Nearly all
possible  scenarios of parton  distributions  proposed have favoured a
more or less gluonic content.  In the following  section we shall give
a very cursory presentation of the most popular Pomeron models at this
stage.  A  pivotal  role  will be  played  by  three  Pomeron  models,
recently  presented and applied to diffractive  $W$  production at the
Tevatron and the LHC by Kunszt and Stirling \cite{james96}.  They will
be  the  input   distributions  for  the  diffractive   heavy  flavour
production, which we shall deal with in Section~4.


\section{Models of the Pomeron}

The Pomeron has been postulated to describe hard diffractive  collider
phenomenology.  In the  framework  of Regge  theory a colourless  {\em
object},  carrying  the  quantum  numbers  of the  vacuum,  is able to
explain the new  observations.  A pure gluonic  composite seemed to be
the  simplest  explanation  on the  basis  of  partonic  contents  and
proposed features  \cite{low75}.  The Pomeron was therefore assumed to
behave essentially as a hadron and Ingelman and Schlein introduced the
concept of a {\em Pomeron structure function} \cite{ingelman85}.

Further collider  experiments, by for example the UA8 collaboration at
CERN \cite{UA8}, gave evidence for a hard parton  distribution  inside
the Pomeron but still could not  distinguish  between a gluon or quark
dominated Pomeron.

Another  striking  problem,  not yet  fully  resolved,  was  the  {\em
character}  of the Pomeron:  should it be treated by  perturbative  or
non--perturbative  means?  Landshoff  \cite{landshoff92}  argued  that
experimental   results  are  best   described  by  a  Pomeron   having
propagators that are  non--perturbative.  They must not have a pole at
$k^2=0$.  He   summarises   the   phenomenological   features  of  the
so--called {\em soft} Pomeron as follows:

\begin{itemize}

\item{ it is rather like a ${\cal{C}}=+1$ isoscalar photon,}

\item{   it   couples    to   single    quarks    with   a    strength
$\beta=1.8$~GeV$^{-1}$,}

\item{it is a simple Regge pole with a Regge trajectory $\alpha(|t|) =
1 + \epsilon -  \alpha'|t|$  taking into  account an  effective  power
$\epsilon$ that decreases with increasing  $\sqrt{s}$, but very slowly
such  that  $   \epsilon\approx   0.08$  at  typical  hadron  collider
energies.}  

\end{itemize}

Studies at the $p\bar{p}$--collider at CERN \cite{UA8} favoured a more
hard quark-- or gluon--like  structure function for the Pomeron than a
soft  one\footnote{The  conclusion  of UA8 was that  $\sim57$\% of all
events should be {\em hard} $\lbrack  6x(1-x)  \rbrack$ and $\sim13$\%
were  tagged  to  be  {\em  soft}  $\lbrack  6(1-x)^5   \rbrack$.  The
remaining 30\% showed a {\em superhard} $\lbrack \delta(1-x)  \rbrack$
characteristic.}.  Even  though the Pomeron  exchange  is a pure gluon
exchange, this does not imply that the Pomeron  structure  function is
entirely, or even predominantly, gluonic.

This was the starting point to give predictions for diffractive events
using  different  Pomeron models, for example for  diffractive $W$ and
$Z$  boson  production   \cite{bruni93}   or  for  diffractive   Higgs
production \cite{nachtmann90}.

Whatever  the  parton  distributions  looked  like,  in  the  original
investigations  they were assumed to show no evolution  with the scale
$Q^2$ at which the Pomeron is probed.  They were functions only of the
momentum  fractions of the partons inside the Pomeron.  This situation
has changed,  since there are now data for the  diffractive  structure
functions measured at HERA  \cite{H194,ZEUS94}  at different values of
$Q^2$.  Employing   Eq.~(\ref{factor1})   and  assuming  factorisation
directly   yields   data   for   the   Pomeron   structure    function
$F^{\rm\pom}_2(x,Q^2)$,  assuming  a  certain  structure  for the flux
factor.  This has been discussed in detail in the Introduction.

We shall focus on the latest set of three  different  Pomeron  models,
based on recent HERA data \cite{HERA}, provided by Kunszt and Stirling
\cite{james96}.  The main features of these models are:

\begin{itemize}

\item{all three models are qualitatively very different concerning the
partonic contents at the starting scale $Q_0=2$~GeV,}

\item{all  three models give  satisfactory  agreement  with the H1 and
ZEUS data,}

\item{all  models  undergo  leading--order  GLAP  evolution as for the
usual parton distributions.}  

\end{itemize}

In qualitative terms, the three models can be characterised as:

\begin{description}

\item[Model  1:]{At $Q_0$ the Pomeron is entirely  composed of quarks.
Gluons are dynamically generated via GLAP evolution;}

\item[Model 2:]{A mix of quarks and gluons at starting scale $Q_0$;}

\item[Model  3:]{A  predominantly  hard  gluonic  content at  starting
scale, the gluons inside the Pomeron carry large fractional  momenta.}

\end{description}

An SU(3) flavour symmetry is assumed for all three models, such that a
momentum sum rule for $Q_0=2$~GeV can be imposed
\beq  
\int\limits_0^1  dx\,  x  \left(  6  f_{q/{\rm\pom}}(x,Q_0^2)  +
f_{g/{\rm\pom}}(x,Q_0^2)\right)  = 1,           
\eeq
to deduce the overall  normalisation factor for the data fit.  Because
the quark contents inside the three models are more or less comparable
since they are  constrained by the HERA data, we show in Fig.~3 solely
the $Q$ evolution of the gluon distributions $xf_{g/{\rm\pom}}(x,Q^2)$
for the three Pomeron models.

We  shall  see  later  that  it  is  the  $gg$   contributions   which
predominantly  govern the  behaviour of the  (diffractive)  $Q\bar{Q}$
cross  sections  for  the  Tevatron  and the LHC in the  given  energy
regime.  These cross  sections  can  therefore  in  principle  provide
important   information   on  the  gluon   content  of  the   Pomeron,
complementary  to the  constraints  on  the  quark  content  from  DIS
structure functions.  For a broader discussion of the partonic Pomeron
contents  and  their  fit  to the H1  data  we  refer  the  reader  to
\cite{james96}.

It  is  very  important  to  have   distributions  that  undergo  GLAP
evolution,  because, ranging from charm to top quark masses, different
$Q^2$   scales  are  probed  and,  as  can  be  seen  in  Fig.~3,  the
distributions evolve significantly with $Q^2$.  This is different from
most of the former {\em static} distributions  proposed and enables us
to give  predictions  for such a large mass range covered by the heavy
flavours.

The  main   features   of  the   three   Pomeron   models   concerning
$xf_{g/{\rm\pom}}  (x,Q^2)$  can easily be deduced  from  Fig.~3.  The
left side shows surface plots of the gluon distributions in the regime
$4$~GeV   $\leq  Q  \leq   50$~GeV  and  $0.1\leq  x  \leq  1.0$.  The
corresponding  contour plots in the  $x$--$Q$  plane  are presented on
the right side.

As  can  be  seen  in  Fig.~3,   Model~1  gains  only  a  small  gluon
distribution  via GLAP  evolution at $Q=4$~GeV.  Recall that there are
no  gluons  at all  at  the  starting  scale  $Q_0=2$~GeV.  The  gluon
distribution  is almost {\em flat} in $Q$ for moderate and high values
of $x$.  This  fact can also be seen in the  contour  plot,  where for
$Q>10$~GeV there is only a significant evolution in the very small $x$
regime.  No gluons with $x>0.6$ are present in this regime.

This is different to Model~2, which  contains  gluons with moderate or
even  higher  fractional  momenta.  The   gluon  distribution  shows a
significant $Q$ evolution not only for small but also for moderate $x$
(this can be seen  clearly  in the  contour  plot),  even  though  the
distribution   asymptotically  approaches  zero  for  higher  $Q$  and
$x>0.3$.

Model~3  finally  shows  the most  dynamical  evolution  of the  gluon
distributions  not only in $Q$ but also in $x$.  Very hard  gluons are
present for high values of $x$.  This is due to the basic construction
of Model~3 as discussed  above.  These  gluons lose their high momenta
at  larger  values  of  $Q$  mainly  due  to   quark--antiquark   pair
production.  The  depletion of these high  frac\-ti\-onal--mo\-men\-ta
gluons with  increasing  $Q$ again  becomes  very  transparent  in the
corresponding  contour plot.  For a further  discussion we again refer
the reader to~\cite{james96}.


\section{Diffraction at the Tevatron and the LHC}

For  the  GLAP  evolution  of  the  parton  distributions  inside  the
(anti)proton  we use the MRS(A$'$) set  presented  in~\cite{martin95}.
One  test of the  reliability  of this  set is the  comparison  of the
calculated total inclusive cross section for $t\bar{t}$  production at
$\sqrt{s}  =  1.8$~TeV  with  the  latest   experimental   measurement
\cite{top95}.  The CDF collaboration  measures a total inclusive cross
section of  $\sigma^T_{exp.}(p  + \bar{p} \rightarrow  t\bar{t} + X) =
7.6^{+2.4}_{-2.0}$~pb for an experimentally determined top mass $m_t =
(\,176\pm  8({\rm  stat.})  \pm  10({\rm  syst.})\,)$~GeV,  while  the
MRS(A$'$) set of partons  predicts  $\sigma^T(p + \bar{p}  \rightarrow
t\bar{t} + X) = 5.81$~pb for $m_t=176$~GeV.  Measurements of inclusive
$c\bar{c}$  and  $b\bar{b}$  cross  sections at the Tevatron  collider
yield~\cite{CDF}:
\begin{eqnarray*}  
\sigma_{\rm exp.}^T( p + \bar{p} \rightarrow c\bar{c} + X) &=& 
115.2^{+24.3}_{-26.8}~{\rm nb},\\ \sigma_{\rm exp.}^T( p + \bar{p}
\rightarrow   b\bar{b}   +  X)   &=&   30.1^{+17.0}_{-17.4}~{\rm   nb}
\end{eqnarray*}
for $m_c = 1.5$~GeV and $m_b = 4.5$~GeV compared to the leading--order
theoretical  predictions of $\sigma^T(p + \bar{p} \rightarrow c\bar{c}
+ X) = 98.7$~nb and $\sigma^T(p + \bar{p} \rightarrow  b\bar{b} + X) =
44.2$~nb  for the same $m_c$ and  $m_b$.  Even  though we are  working
only  in   leading--order   approximation,   the  agreement  with  the
experimental  results  is  encouraging.  Predictions  for  the  ratios
$\sigma^{SD}/\sigma^{T}$ and $\sigma^{DD}/\sigma^{T}$, however, should
be relatively insensitive to higher--order perturbative corrections.

\subsection{Probing the Pomeron at the Tevatron}

We calculate  numerically the total inclusive, the single  diffractive
and  the  double   diffractive   cross   sections  for  heavy  flavour
pair--production  using the  MRS(A$'$)  set of  partons  and the three
Pomeron  models   introduced  in   \cite{james96}   and  discussed  in
Section~3.  Our  results  are shown in Fig.~4.  Supplementary  we show
the pure gluon fusion contribution as dashed lines.

For the Tevatron collider the $q\bar{q}$  process becomes dominant for
$m_Q>50$~GeV.  This $q\bar{q}$  dominance is obviously  visible in the
total cross  section  but becomes  even more  striking  for the single
diffractive  one.  Due to this process, the single  diffractive  ratio
$R^{SD}$   defined   as   $R^{SD}=\sigma^{SD}/\sigma^{T}$   even  {\em
increases}   for   $m_Q>70$~GeV   before  this  process  runs  out  of
centre--of--mass  energy  $\sqrt{\hat{s}}$  due to the cut--off in the
Pomeron spectrum  ($x_i^{\rm\pom}\leq0.1$)  and steeply falls to zero.
Note that this  restriction on  $\sqrt{\hat{s}}$  unfortunately  takes
place before the top quark  production  domain is reached.  The single
diffractive  ratio  $R^{SD}$ for all three Pomeron  models reaches its
local maximum at slightly  lower masses than the top quark mass.  This
gives  little  hope to  observe  diffractive  top quark  events at the
Tevatron  but is  promising  for  the  LHC as  will  be  discussed  in
Section~4.2.

At Tevatron  energy the  threshold  for single  diffractive  events is
reached when  $m_Q=270$~GeV.  But single  diffractive  events might be
{\em observed} only up to $m_Q\approx150$~GeV  with
\begin{eqnarray*}   
\label{single}  
{\rm\bf  Model}\;{\rm\bf  3:}\quad
\sigma_{\rm  max.}^{SD}  &=&  1.05~{\rm  pb}  
\simeq  4.05\%\;\;  {\rm of}\;\; \sigma^T, \\
{\rm\bf Model}\;{\rm\bf  1:}\quad 
\sigma_{\rm min.}^{SD} &=& 0.22~{\rm pb}  \simeq   0.86\%\;\;   
{\rm   of}\;\;   \sigma^T   
\end{eqnarray*}
for $m_Q=150$~GeV.    This is  qualitatively  what we expect.  Model~3
with the hard gluon content gains more and more  $q\bar{q}$  pairs via
GLAP evolution which dominantly  contribute to the single  diffractive
cross  section  for  higher  masses.  Model~2  with  the  quark--gluon
mixture dominates for small masses, but at $m_Q\simeq  10$~GeV Model~3
takes over.  The production rate of the important $q\bar{q}$ pairs for
higher masses  increases  more rapidly in Model~3.  Model~1,  however,
without any gluons at the initial  $Q_0=2$~GeV scale,  marks the lower
limit for single and double diffractive scattering over the whole mass
range.  The  numerical  values for the  single and double  diffractive
production of charm, bottom and top are given in Table~1.

Let us now focus on Models~2 and 3:  the ratios for single diffractive
scattering  range  between  $\sim 22\%$ for charm and $\sim  15\%$ for
bottom quark production and reach their locally lowest values of $\sim
2\%$ for  $m_Q=70$~GeV,  before the ratio  even  increases  due to the
$q\bar{q}$  contribution  in the mass region  $70$~GeV  $\leq m_Q \leq
150$~GeV locally peaking at $m_Q = 150$~GeV with ratio $R^{SD} = 5.5\%
(4.2\%)$ for Model~3~(2).  {\em Qualitatively}  Model~1 shows the same
behaviour.  But with only  quarks in the  starting  distribution,  the
creation of gluons and  $q\bar{q}$  pairs via GLAP evolution  proceeds
only slowly.  This can be seen in Fig.~3.  The difference is expressed
in the large gap between Model~1 on the one hand and Models~2 and 3 on
the  other  hand.  So the  conclusion  is  that at  least  for  single
diffraction  a gluon  rich  Pomeron  as input  should  be more  easily
detected.  Model~1 and Model~3 for  example  differ by about one order
of  magnitude.  This  turns out to be a crucial  difference  with such
absolutely small diffractive cross sections.

Double  diffractive  scattering seems to favour a balanced  mixture of
quarks  and  gluons  in the  starting  distribution  and  during  GLAP
evolution,  as  provided  in Model~2.  This can again be  observed  in
Fig.~4.

Notice   that  the   former   dominant   hard   gluon   Model~3   {\em
quantitatively}  shows about the same  behaviour  as Model~1.  Why the
behaviour of Model~3 in double  diffraction  is different from that in
single  diffraction  can be  understood  by a further  analysis of the
kinematics   among  the   partons   inside  the   Pomeron   and  their
distributions  at  different  mass scales.  Models~2 and 3 show a {\em
crossing}  in the  case of  single  diffractive  scattering  as can be
observed in Fig.~4.  Its  existence  can be  immediately  explained in
terms of the gluon  distributions  which are  shown in  Fig.~3.  As we
have already pointed out, the gluon distributions govern the behaviour
of the cross  sections  at the  Tevatron,  especially  for small quark
masses.  For small $x$ and $Q$, the gluon  distribution  of Model~2 is
slightly  bigger than that of  Model~3.  For higher $x$ and/or $Q$ the
situation is reversed.

In the  case  of  double  diffraction,  an  analysis  of  the  average
fractional    gluon   momentum    $\langle   x   \rangle   =   \langle
x_i/x_i^{\rm\pom}  \rangle$  yields  $\langle x \rangle \sim 0.11$ for
$m_Q=2$~GeV and $\langle x \rangle\sim  0.24$ for $m_Q=50$~GeV for all
three models.  In this regime the gluon  distribution of Model~2 again
exceeds that of Model~3.  So, no crossing  can be observed and Model~2
dominates  throughout.  In fact  the  crossing  would  take  place  at
$m_Q\sim 80$~GeV, shortly before the threshold for double  diffraction
with $m_Q\sim 90$~GeV is reached.

The  relatively  broad gap between  Model~2 and Model~3  appearing  in
Fig.~4 can again be explained by the same straightforward  analysis of
the average gluon momentum.  Again a comparison with the corresponding
gluon  distributions  of  Model~2  and  Model~3  in  Fig.~3  shows the
absolute   difference   of   these   distributions   in   the   regime
$0.1\leq\langle x \rangle\leq0.2$.

While   diffractive   charm  and  bottom  quark  production  might  be
observable  at  the  Tevatron,  there  is  no  hope  for   diffractive
$t\bar{t}$ pair production  being visible at a total  centre--of--mass
energy  of   $\sqrt{s}=1.8$~TeV.  Either  the  top  mass  exceeds  the
kinematic  threshold  (double   diffraction)  or  the  effect  of  the
threshold  is already  strongly  influencing  the  process  by a steep
decrease in the diffractive cross section (single diffraction near top
mass).  However at the LHC such heavy flavour threshold suppression is
less severe, as we shall see in the following Section.

\subsection{Predictions for the LHC}

The LHC will  provide a rich  field of study for  diffractive  events,
even in the top mass  regime.  With a  centre--of--mass  energy  of at
least $\sqrt{s} = 10$~TeV \cite{LHC} the double diffractive  threshold
is reached for  $m_Q=500$~GeV  and the single  diffractive one lies at
$m_Q=1500$~GeV.  This upper bound is very promising in particular  for
diffractive top quark production at the LHC.

At a  proton--proton  collider,  the dominant  process for  $Q\bar{Q}$
production  is of course  gluon--gluon  fusion since  antiquarks  only
appear as sea quarks  inside the proton.  For a top mass of  $176$~GeV
the pure $gg$  contribution is about $91\%$ of the total cross section
in our  calculation.  This  also  holds  for  the  single  and  double
diffractive case.

The numerical results are shown in Fig.~5 and the numerical values are
again listed in Table~1.  The single  diffractive  ratios $R^{SD}$ for
the three Pomeron models are between $20-40\%$ for charm and $10-40\%$
for bottom  quarks.  The maximal and minimal  single  diffractive  top
quark rates are
\begin{eqnarray*}   
{\rm\bf   Model}\;{\rm\bf   3:}\quad   \sigma_{\rm
max.}^{SD} &=& 5.45~{\rm pb} \simeq 1.66\%\;\; {\rm of}\;\;  \sigma^T,
\\
{\rm\bf Model}\;{\rm\bf  1:}\quad \sigma_{\rm min.}^{SD} &=& 1.13~{\rm
pb}  \simeq   0.34\%\;\;   {\rm   of}\;\;   \sigma^T   
\end{eqnarray*}
with $m_t=176$~GeV.  Even though the single diffractive ratios for top
production are  comparable to the Tevatron  rates, the {\em  absolute}
single diffractive cross sections are crucially  enhanced.  We find an
enhancement of about a factor 100 at the LHC.  For example Model~2 and
Model~3 provide a single  diffractive  cross section of  approximately
$5$~pb.  This  is  about  the  {\rm  total}  cross   section  for  top
production at the Tevatron.  So, single  diffractive  top quark events
should be readily detected at the LHC.

The predictions for double diffractive  scattering, however, are still
not very promising.  The maximal double diffractive cross section (for
Model~2) is $1.86\cdot  10^{-2}$~pb  for top quarks.  The  qualitative
behaviour of the three different  Pomeron models is the same.  Because
the  centre--of--mass  energy at the LHC is larger by a factor of six,
one can conclude  from the  qualitatively  similiar  behaviour  of all
three  models  at the LHC that the gluon  distributions  in all  three
models do indeed  become  similiar  at higher $Q$, as already  seen in
Fig.~3.

All  conclusions  that were drawn for the Tevatron  still hold for the
LHC, except that because of the higher  centre--of--mass  energy there
are no  kinematical  artifacts in the considered  flavour mass regime.
Even  for the top  quark  mass,  the  single  as  well  as the  double
diffractive  cross section  behave rather  smoothly.  The influence of
the  threshold  does not seriously  affect the cross  sections in this
case.

Considering  the ratios for single and  double  diffractive  events in
Fig.~5 again shows a qualitatively comparable picture to the Tevatron.
Again an analysis of the average  fractional gluon momentum $\langle x
\rangle$ will explain the  differences  in $R^{SD}$ and $R^{DD}$.  For
$R^{SD}$,  Models~2  and  3  are  {\em   quantitatively}   equivalent,
especially  for the charm and bottom  quarks.  For a bottom quark mass
of $m_b = 4.5$~GeV,  we obtain as average  fractional  gluon  momentum
$\langle  x  \rangle = 0.42$  for  $Q\sim  2m_b=9$~GeV  in the  single
diffractive case.  A comparison with the gluon distributions in Fig.~3
shows that they are  roughly  equal for  Model~2  and  Model~3 in this
region of $x$.  This fact becomes even more transparent in the contour
plots of Fig.~3.

For the case of double  diffraction  the same analysis  yields a lower
average  fractional  momentum  for the gluons, due to the  energy--cut
among  both  Pomeron  emitting  hadrons.  Again for  $m_b=4.5$~GeV  we
obtain  $\langle  x \rangle = 0.22$.  But in this  region  of $x$, the
gluon  distribution  of Model~3 shows a local minimum, the hard gluons
in  this  model  give a rise  of  $xf_{g/{\rm\pom}}(x,Q^2)$  only  for
$x>0.5$ in the low--Q regime.  The gluon distributions  inside Model~2
also show a local minimum  around $x\sim 0.2$, but its absolute  value
is  higher  than  that  for  Model~3  in  this  region.  This  fact is
responsible  for the gap between  Model~2 and Model~3 as observed  for
$R^{DD}$ in Fig.~5.  For higher quark masses  (higher values of $Q^2$)
all three models become comparable concerning the gluon distributions,
as already discussed.

Model~2 and Model~3 as descriptions of the parton distributions of the
Pomeron yield very promising single  diffractive  ratios, at least for
charm and bottom quarks.  Model~3 predicts a single  diffractive ratio
of $\sim  40$\% for  $c\bar{c}$  and  $b\bar{b}$  production.  This is
quantitatively  comparable  to the  predictions  of Model~2  as can be
observed in Fig.~5 and  numerically  verified  in Table~1.  Thus about
one third of the  production  of heavy  flavours at the LHC  including
charm  and  bottom  quarks  should  be  single  diffractive.  But even
Model~1,  purely  quark--like at starting scale  $Q_0=2$~GeV,  gives a
single  diffractive  contribution of  approximately  20\% in this mass
regime.


\section{Conclusions}

We have presented predictions for diffractive heavy flavour production
at  the  Tevatron  and  the  LHC.  The  input  Pomeron   distributions
\cite{james96}  were  different in contents  from pure quark  starting
distributions  up to hard gluonic  contents at the GLAP starting scale
$Q_0=2$~GeV.  Our predictions can therefore be considered as upper and
lower limits for the diffractive cross sections.

We found that at the Tevatron single and double  diffractive charm and
bottom quark production is observable with a single  diffractive ratio
$R^{SD}$ that is between $7\%$ (charm) and $4\%$  (bottom) for Model~1
as a lower limit and $26\%$ (charm) and $13\%$ (bottom) for Model~2 as
an upper bound.  The double  diffractive  cross  section for charm and
bottom production lies above the {\rm total} cross section for the top
quark and might also be  observable.  The lower limit is  governed  by
the pure quark--like  Model~1 with  $\sigma^{DD}\sim  45$~pb for charm
and $\sigma^{DD}\sim  18$~pb for bottom quarks, whereas having Model~2
as the  underlying  Pomeron  model  gives a double  diffractive  cross
section which is enhanced by a factor of four.

The  diffractive  cross  sections for charm and bottom  quarks will be
larger  by three  orders  of  magnitude  at the LHC,  and also  single
diffractive  top quark  production  should be  observable.  The single
diffractive $t\bar{t}$ cross sections range from $1$~pb for Model~1 up
to $5$~pb for the hard gluon  Model~3,  which is roughly the total top
quark cross section at the Tevatron.  All predictions are collected in
Table~1 and shown in Figs.~4 and 5.

The three Pomeron models presented in  \cite{james96}  and used in our
calculations  are only a guide to possible Pomeron  structures.  Their
determinations  have been made  possible  by two  strong  assumptions:
first, the  Pomeron  has a partonic  structure  and  second, we assume
factorisation such that recent measurements of the diffractive Pomeron
structure  function  give a {\em  universal}  picture of the  Pomeron.
Only the latter  makes  predictions  for the  hadron--hadron  collider
meaningful.

Our  predictions  {\em  can} only be a guide as the  calculations  are
illustrative  only.  We  need  a more  detailed  study  with  detector
acceptance.  This will  follow as soon as there are first  results  on
diffractive heavy flavour  production at the Tevatron.  But a study of
these events {\em is}  worthwhile, at least for diffractive  charm and
bottom  production.  Diffractive top quark events still will require a
higher  centre--of--mass  energy  before they can be studied.  The LHC
will be such a  laboratory  for  diffractive  scattering  studies.  We
predict that about  $30\%$ of all charm and bottom  production  events
should be single diffractive.

It is not  yet  certain  whether  the  theoretical  framework  of  the
Donnachie--Landshoff  model can be extended to double or even multiple
Pomeron  exchange.  Again,  we give  predictions  for  the  case of an
analogue treatment of double Pomeron exchange that will be interesting
to test at future hadron--hadron  colliders.  Also the parameters used
in this  theoretical  model  have  to be  fitted  to new  high  energy
collider experiments\footnote{Especially we expect the effective power
$\epsilon$  inside  the  Pomeron  flux  factor to become  smaller  for
typical LHC energies.}.  Finally future high energy  experiments  will
hopefully  more clearly reveal the structure of the Pomeron and answer
some  questions  that  were  raised  in this  paper.


\section*{Acknowledgements}

I would like to thank James  Stirling for  suggesting  the problem and
many  fruitful discussions  throughout  this work.  John  Campbell  is
thanked for a critical reading of this manuscript.  Financial  support
in  the  form  of a  ``DAAD--Doktorandenstipendium''  HSP--II/AUFE  is
gratefully acknowledged.



\newpage


\begin{table}[h]

\begin{tabular}               {|l||c|c|c||c|c|c|}\hline              &
\multicolumn{3}{|c||}{\shift   Tevatron   ($\sqrt{s}$  =  1.8~TeV)}  &
\multicolumn{3}{   c|}  {\shift   LHC   ($\sqrt{s}$   =   10.0~TeV)}\\
\cline{2-7}\hline   &   \multicolumn{3}{|c||}{\shift    $\langle   m_c
\rangle$}  &  \multicolumn{3}{  c|}{\shift  $\langle  m_c  \rangle$}\\
\cline{2-7}  {\shift model} & 1 & 2 & 3 & 1 & 2 & 3 \\ \hline  {\shift
$\sigma^{SD}   \lbrack  {\rm  pb}  \rbrack$}  &  $7.81\cdot   10^3$  &
$29.82\cdot 10^3$ & $24.32\cdot 10^3$ & $2.07\cdot 10^6$ & $ 3.59\cdot
10^6$ & $  3.78\cdot  10^6$  \\  \hline  {\shift  $R^{SD}  \lbrack  \%
\rbrack$}  & $6.83$ & $26.15$ & $21.27$ & $21.43$ &  $37.21$ & $39.21$
\\ \hline {\shift  $\sigma^{DD} \lbrack {\rm pb} \rbrack$} & $45.32$ &
$208.12$ & $60.68$ & $4.05\cdot  10^4$ & $6.56\cdot 10^4$ & $4.44\cdot
10^4$ \\ \hline  {\shift  $R^{DD}  \lbrack \%  \rbrack$} &  $3.96\cdot
10^{-2}$ &  $18.21\cdot  10^{-2}$  &  $5.31\cdot  10^{-2}$  & $0.42$ &
$0.68$          &           $0.46$           \\           \hline\hline
&    \multicolumn{3}{|c||}{\shift    $\langle    m_b    \rangle$}    &
\multicolumn{3}{   c|}{\shift  $\langle  m_b  \rangle$}\\  \cline{2-7}
{\shift model} & 1 & 2 & 3 & 1 & 2 & 3 \\ \hline {\shift  $\sigma^{SD}
\lbrack {\rm pb} \rbrack$} & $3.07\cdot  10^3$ &  $11.59\cdot  10^3$ &
$10.68\cdot 10^3$ & $4.50\cdot 10^5$ & $ 1.01\cdot 10^6$ & $ 1.11\cdot
10^6$ \\  \hline  {\shift  $R^{SD}  \lbrack  \%  \rbrack$}  & $3.31$ &
$12.49$ & $11.51$ & $15.50$ & $34.53$  &  $38.27$  \\  \hline  {\shift
$\sigma^{DD}  \lbrack {\rm pb} \rbrack$} & $17.82$ & $78.85$ & $25.26$
&  $5.81\cdot  10^3$ & $1.51\cdot  10^4$ & $7.55\cdot  10^3$ \\ \hline
{\shift   $R^{DD}  \lbrack  \%  \rbrack$}  &  $1.92\cdot   10^{-2}$  &
$8.49\cdot  10^{-2}$ & $2.72\cdot  10^{-2}$ & $0.20$ & $0.52$ & $0.26$
\\                                                        \hline\hline
&    \multicolumn{3}{|c||}{\shift    $\langle    m_t    \rangle$}    &
\multicolumn{3}{   c|}{\shift  $\langle  m_t  \rangle$}\\  \cline{2-7}
{\shift model} & 1 & 2 & 3 & 1 & 2 & 3 \\ \hline {\shift  $\sigma^{SD}
\lbrack {\rm pb} \rbrack$} & $1.86\cdot 10^{-2}$ & $6.06\cdot 10^{-2}$
&  $8.72\cdot  10^{-2}$  & $1.13$ & $4.10$ & $5.45$ \\ \hline  {\shift
$R^{SD}  \lbrack  \%  \rbrack$}  & $0.31$ & $1.01$ & $1.45$ & $0.34$ &
$1.25$ & $1.66$  \\  \hline  {\shift  $\sigma^{DD}  \lbrack  {\rm  pb}
\rbrack$} & $\bigcirc$ & $\bigcirc$ & $\bigcirc$ & $1.04\cdot 10^{-4}$
& $1.86\cdot 10^{-2}$ & $3.71\cdot  10^{-3}$ \\ \hline {\shift $R^{DD}
\lbrack  \%  \rbrack$}  &  $\bigcirc$  &  $\bigcirc$  &  $\bigcirc$  &
$3.17\cdot  10^{-4}$ & $5.68\cdot  10^{-3}$ &  $1.13\cdot  10^{-3}$ \\
\hline \end{tabular}

\caption{The  values for single and double diffractive cross sections,
as well as their  ratios to the total  cross  sections,  are shown for
average quark masses $\langle m_c \rangle = 1.3~{\rm  GeV}$,  $\langle
m_b  \rangle = 4.3~{\rm  GeV}$ and  $\langle  m_t  \rangle =  176~{\rm
GeV}$.  The three different Pomeron models are discussed in Section~3.
We  obtain  numerical  data for  both  the  Tevatron  and the  LHC.  A
$\bigcirc$   indicates   that  the  threshold  for  this  process  was
exceeded.}

\end{table}

\newpage


\begin{figure}[b] \unitlength1cm \begin{center}  \begin{picture}(14,8)
\makebox[14cm]{\epsfxsize=14cm  \epsfysize=8cm  \epsffile{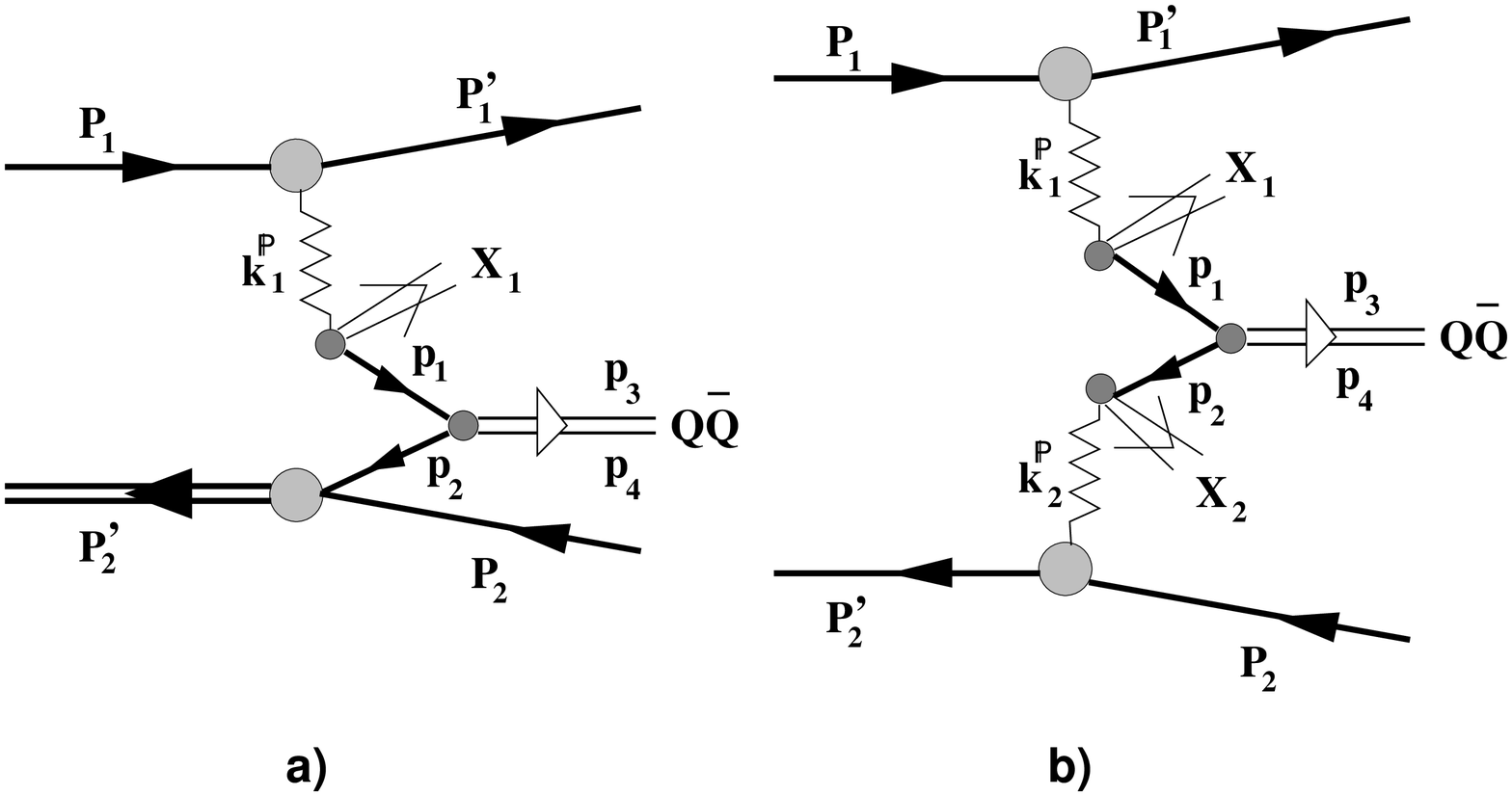}  }
\end{picture} \end{center}

\caption{Kinematics of the single {\bf (a)} and the double diffractive
scattering  {\bf (b)} events,  leading to $Q\bar{Q}$ pair  production.
The subprocesses  taken into account are either $q\bar{q}  \rightarrow
Q\bar{Q}$ or $gg \rightarrow Q\bar{Q}$ .  }

\end{figure}


\begin{figure}[b] \unitlength1cm \begin{center}  \begin{picture}(14,8)
\centering        \makebox[14cm]{\epsfxsize=14cm        \epsfysize=8cm
\epsffile{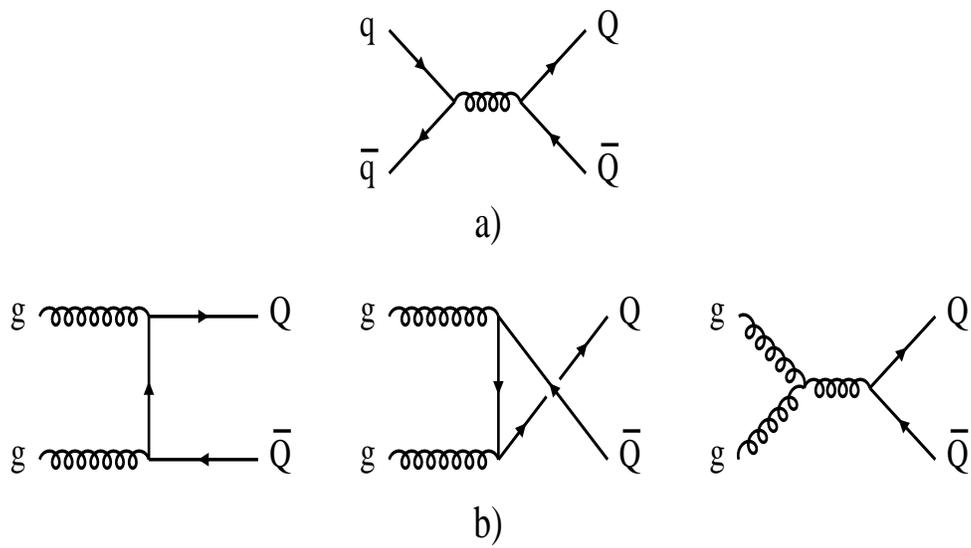} } \end{picture} \end{center}

\caption{ The  leading--order  Feynman  diagrams for the  subprocesses
$q\bar{q}   \rightarrow   Q\bar{Q}$  {\bf  (a)}  and  $gg  \rightarrow
Q\bar{Q}$ {\bf (b)}.}

\end{figure}

\newpage


\begin{figure}[b]             \unitlength1cm            \begin{center}
\begin{picture}(13.2,16) \centering \makebox[13.2cm]{\epsfxsize=13.2cm
\epsfysize=14cm \epsffile{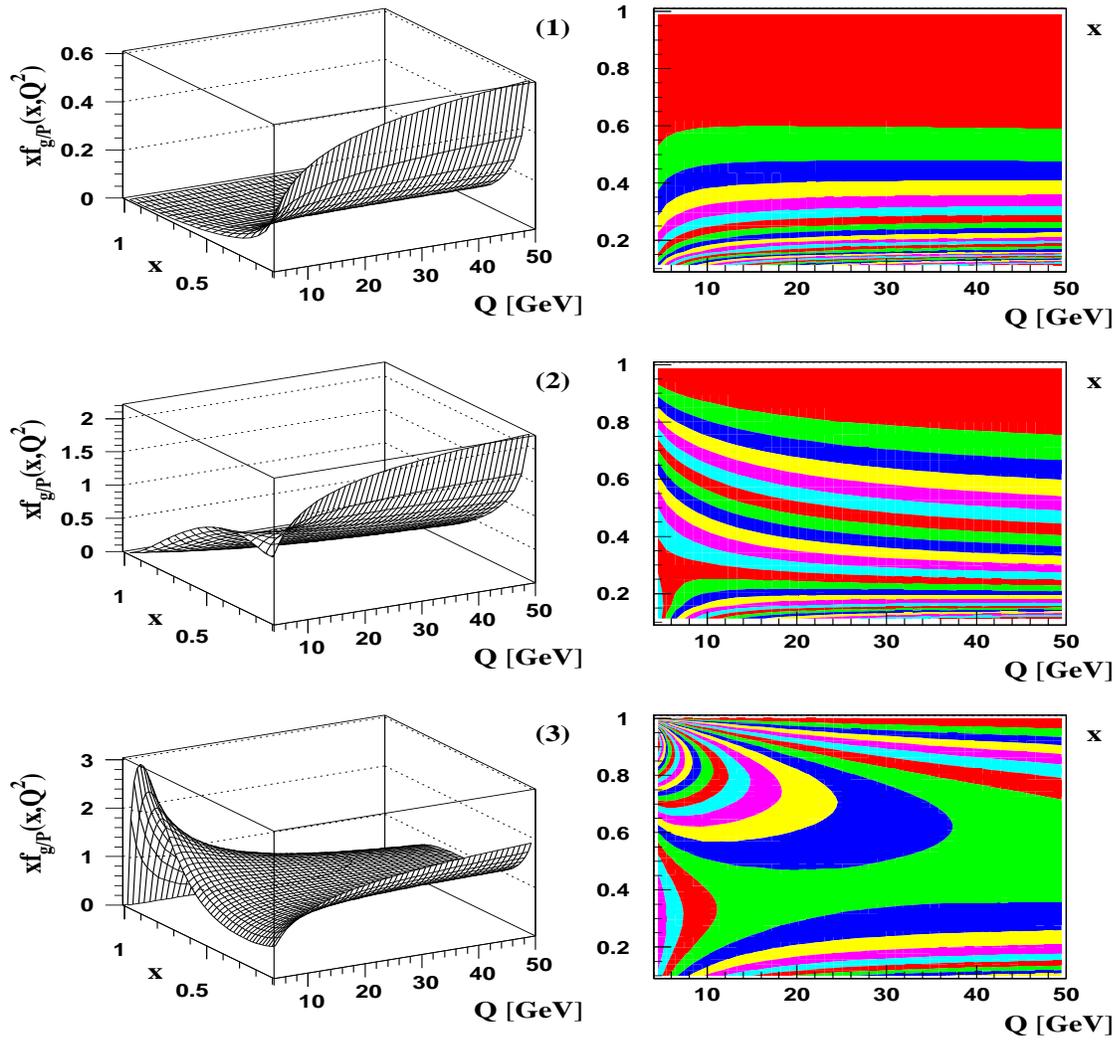} } \end{picture} \end{center}

\caption{The gluon distributions of the three different Pomeron models
are shown,  starting  with Model~1 at the top.  For each model we show
the  surface  plot  for  a  given  $x$--$Q$  regime  as  well  as  the
corresponding contour plot in the $x$--$Q$ plane.}

\end{figure}

\newpage


\begin{figure}[b]             \unitlength1cm            \begin{center}
\begin{picture}(13.2,16)            \makebox[13.2cm]{\epsfxsize=13.2cm
\epsfysize=14cm \epsffile{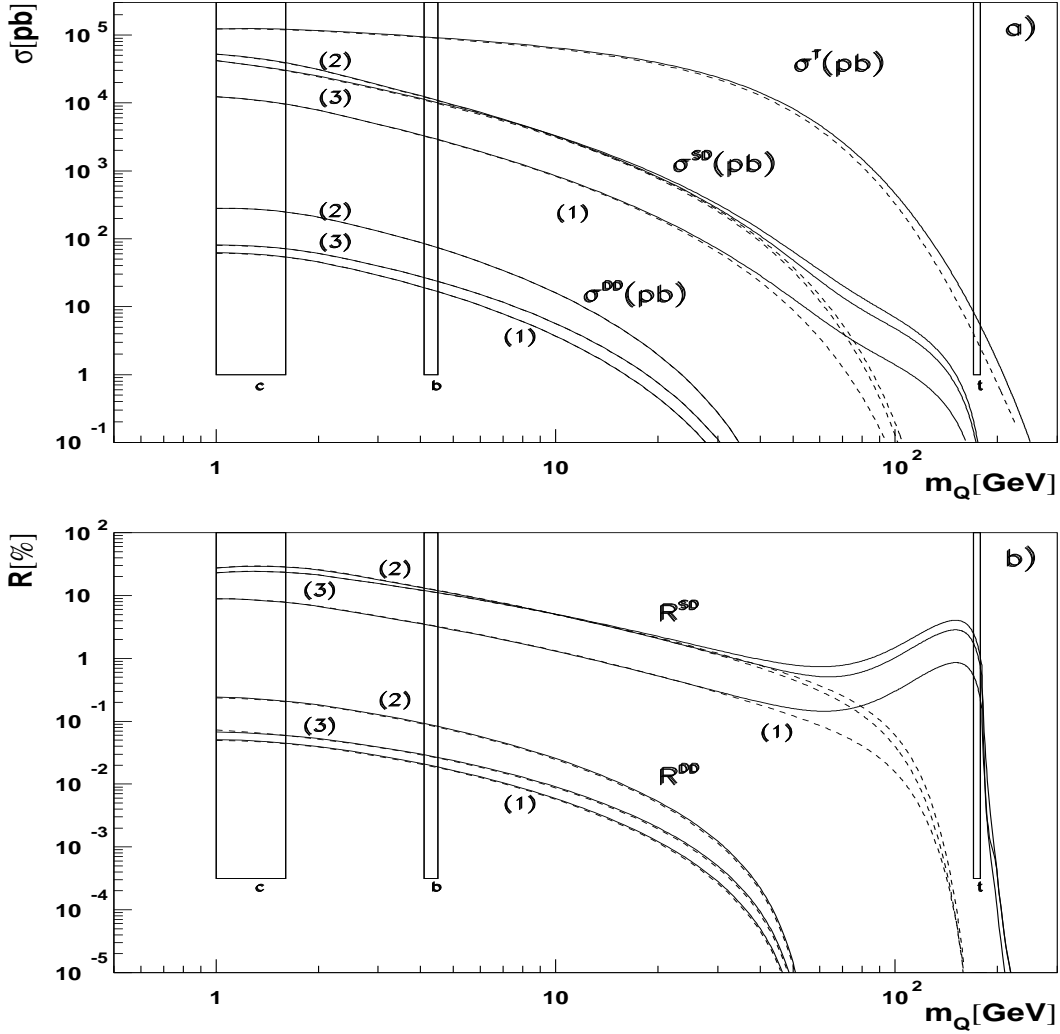} } \end{picture} \end{center}

\caption{ {\bf Diffractive Scattering at the Tevatron:}  The numerical
results of the total, single and double diffractive cross sections for
the three  Pomeron  models are shown.  The solid  lines  indicate  the
subprocess $gg + q\bar{q} \rightarrow Q\bar{Q}$, the dashed lines show
solely the contribution from gluon fusion ($gg \rightarrow Q\bar{Q}$):
{\bf (a)} gives the absolute  numbers for the cross  sections (in pb),
and {\bf (b)} gives the ratios ($R^{SD} = \sigma^{SD}/\sigma^{T}$  and
$R^{DD}  =  \sigma^{DD}/\sigma^{T}$).  The mass  regions  of the charm
(c),  bottom  (b)  and  top  (t)  quarks  are   indicated.  The  fixed
centre--of--mass energy is $1.8$~TeV.}  \end{figure}

\newpage


\begin{figure}[b]             \unitlength1cm            \begin{center}
\begin{picture}(13.2,16)            \makebox[13.2cm]{\epsfxsize=13.2cm
\epsfysize=14cm \epsffile{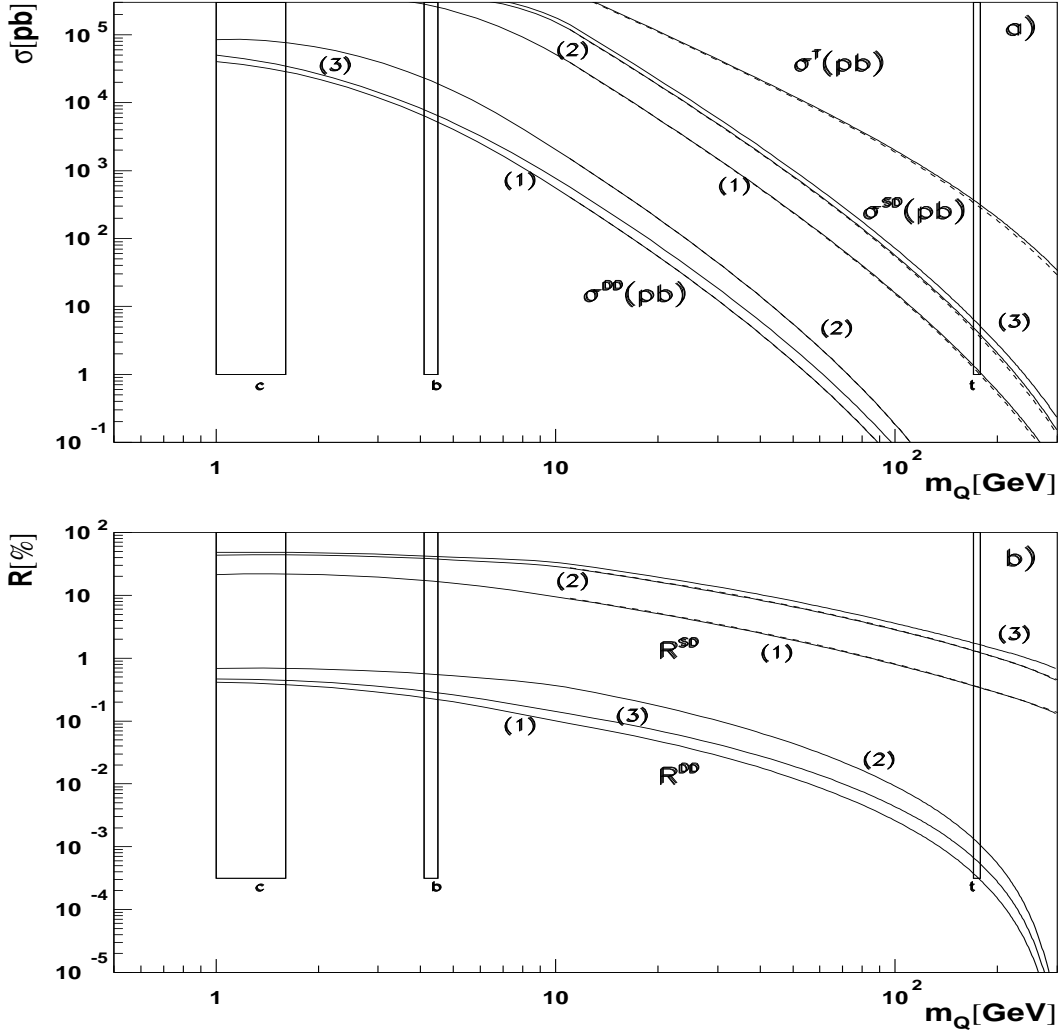} } \end{picture} \end{center}

\caption{{\bf  Diffractive Scattering at the LHC:}  Same as Fig.~4 but
now  for a  centre--of--mass  energy  of  $10$~TeV.  The  solid  lines
indicate the  subprocess  $gg + q\bar{q}  \rightarrow  Q\bar{Q}$,  the
dashed  lines show  solely the  contribution  from gluon  fusion  ($gg
\rightarrow  Q\bar{Q}$): {\bf (a)} gives the absolute  numbers for the
cross  sections  (in pb), and {\bf (b)}  gives the  ratios  ($R^{SD} =
\sigma^{SD}/\sigma^{T}$   and  $R^{DD}  =   \sigma^{DD}/\sigma^{T}$).}
\end{figure}

\end{document}